\newcommand{\be}{\begin{equation}}
\newcommand{\ee}{\end{equation}}
\newcommand{\bea}{\begin{eqnarray}}
\newcommand{\eea}{\end{eqnarray}}
\newcommand{\bi}{\bigskip}

\newcommand{\noi}{\noindent}
\newcommand{\ii}{\'{\i}}
\def\sqr#1#2{{\vcenter{\vbox{\hrule height.#2pt
        \hbox{\vrule width.#2pt height#1pt \kern#1pt
           \vrule width.#2pt}
        \hrule height.#2pt}}}}
\def\square{\mathchoice\sqr54\sqr54\sqr{6.1}3\sqr{1.5}6}

\documentstyle[12pt]{article}
\renewcommand{\thesection}{\Roman{section}}

\begin{document}

\title{\sf Classical solutions in five dimensional induced matter theory
	and its relation to an imperfect fluid}

\author{\sc J. Socorro$^a$, \and V.M. Villanueva$^a$, \and Luis O. 
Pimentel$^b$ }
\date{}
\maketitle

\begin{center}
{$^a${\it Instituto de F\ii sica de la Universidad de Guanajuato} \\
{\it P.O. Box E-143, 37150, Le\'on, Gto., MEXICO}\\
$^b${\it Depto. de F\ii sica, Universidad Aut\'onoma 
Metropolitana-Iztapalapa}\\
{\it P.O. Box 55-534, 09340, D.F., M\'exico, MEXICO.}}
\end{center}
\maketitle
\begin{abstract}
We study five dimensional cosmological models with four dimensional
hipersufaces of the Bianchi type I and V. In this way the
five dimensional vacuum field equations $\rm G_{AB} = 0$,
led us to four dimensional matter equations
$\rm G_{\mu\nu}=T_{\mu\nu}$ and the matter is interpreted as a purely
geometrical property of a fifth dimension. Also, we find that the 
energy-momentum tensor induced from the fifth dimension has the structure
of an imperfect fluid that has dissipative terms.
\end{abstract}
\noi Pacs: {04.50.+h, 98.80.Hw, 02.30.Hq}

\rightline{accepted for publication in IJMP-A}
\rightline{IFUG-JPV1-96}
\rightline{May 7, 1996}

\def\square{\kern1pt\vbox{\hrule height 1.2pt\hbox{\vrule width 1.2pt\hskip
3pt \vbox{\vskip 6pt}\hskip 3pt\vrule width 0.6pt}\hrule height 0.6pt}
\kern1pt}

\renewcommand{\thesection}{\arabic{section}}
\renewcommand{\theequation}{\thesection.\arabic{equation}}

\newpage
\section{Introduction}
\setcounter{equation}{0}
\label{intro}

The problem of higher dimensional cosmologies has received much atention
 as one attempt  to unify interactions.  Most of the people who has worked
this kind of cosmologies has done use of the standard dimensional reduction
of the  Kaluza-Klein type$^1$. A crucial point for the theory to be
realistic is to explain why the ``internal'' space is so small that one
cannot observe it, and why the ``ordinary'' space is so large. An interesting
attempt to this problem is the so called cosmological dimensional
reduction, according to this reduction the large discrepancy of two sizes is
regarded as a consequence of the dynamical evolution of the
higher-dimensional universe.
Nevertheless there have been attempts in which the extra dimensions are not
required to be compact$^{2-4}$, but they are associated to the
geometrical
four-dimensional properties of the matter$^5$ by means of scalar
fields$^6$.

The problem of generating an effective four-dimensional stress-energy tensor
from five-dimensional vacuum field equations (Induced Matter Theory) is not
solved, but we take the ideas of Davidson and Owen$^3$,
Ponce de Leon and
Wesson$^{7,8}$ ({see references in [1], ), in this
way, the four-dimensional
field equations, with their respective stress-energy tensor
$G_{\mu\nu} = T_{\mu\nu},$ can be obtained, there is, of course, an extra
field equation due to the presence of the fifth dimension, namely
$\rm G_{44}$, but this equation is functionally equivalent to
$\rm G^0_0 + G^i_i=0 , i=1,2,3$ or $\rm \rho_{eff}-3p_{eff} =0$ and
$ T_{\mu\nu}$ is a function of just the scale factor of the extra dimension.
The idea of an effective stress-energy tensor in the three-space due to the
evolution of the extra dimensions has been considered by other 
authors$^{9-14}$.


In this work we analyze a specific class of five-dimensional
metrics, that have the following form,

\begin{equation}
\rm ds^2=g_{AB}dx^A dx^B=g_{\mu \nu}(x^\alpha) dx^\mu dx^\nu
+\phi^2(x^\alpha) dw^2,
\end{equation}
where the capital latin indices take values on the five dimensions and
the greek indices take their values on the four dimensions of ordinary
spacetime. Notice that none of the metric coefficients depends on the
fifth coordinate, that is we assume the usual cylinder condition.
In particular, $\rm \sqrt {g_{44}}=\phi$, plays the role of the radius of the
``internal'' space.

Together with the above metric we assume that the action for the theory
of gravity in five dimensions has the following form
\begin{equation}
{\rm I= \int dx^5 \,\sqrt {-^{(5)}g}\, ^{(5)}R. }
\end{equation}
The  Einstein equations from the above action are
\begin{eqnarray}
{\rm ^{(4)}G_{\mu \nu}} &=&{\rm ^{(4)}T_{\mu \nu}(\phi)}:=
{\rm \phi^{-1} \Big[ \phi_{;\mu;\nu}-g_{\mu \nu}
\square \phi \Big],} \nonumber\\
{\rm ^{(5)}G_{44}} &=&{\rm ^{(4)}R} = 0,
\label {FE}
\end{eqnarray}
we will apply these field equations to spaces that are trivial generalizations
of Bianchi  type I and V in the sense that the four dimensional part of
the metric $\rm g_{\mu\nu}$ are of the Bianchi types I and V. For these two
particular examples it is possible to isolate terms that come from the
fifth dimension and associate them with
the effective ``density'' $\rho_{eff}$, and the  effective ``pressure''
 $\rm p_{eff}$ that appear in the stress-energy tensor of a perfect fluid in
four-dimensions. These models have been studied in other contexts, by using
supergravity$^{15,16}$ N=2, D=5. In those models, they found exact
solutions and the existence of singularities is considered.
 In Sec. 2 we present the exact solution for the Bianchi type I, and in
Sec. 3
the solution for the Bianchi type V. In Sec. 4 we obtain the stress-energy
tensor associated with the general models when we  assume the above
decomposition of the metric and  reduce it to four dimensions. It turns out
that in the more general case
the effective energy-momentum tensor has the structure of an imperfect
fluid under some assumptions as have been found in other theories with
a scalar field. In this case, along with besides of the energy density and
the pressure
we have now viscosity, heat flux and anisotropic stress.
This thermodynamic quantities can be useful to explain several features of
 the evolution of the universe. Since
dissipative effects can counteract the collapse a different scenario of the
early stages of the universe appears. Also the generation of entropy can be
accounted for by the dissipative processes. The above identification can
be useful in the following way: to generate new solution in general
relativity with an imperfect fluid from a solution in the theory that we
are considering here.

Some exact solutions with viscosity
have been found for isotropic and anisotropic models, for instance, 
Murphy$^{17}$ found  the solution for flat Friedmann model and Banerjee and
Santos$^{18}$ extended the analysis to the case with curvature.
Belinskii and Khalatnikov$^{19,20}$ have considered the
qualitative aspects in anisotropic Bianchi type-I models and 
Banerjee$^{21,22}$ et al. gave some exact solutions for the Bianchi type-I 
and II models. Bradley and Sviestins$^{23}$ have found exact solution of
Bianchi type VIII with heat flow, while Nayak and Sahoo$^{24}$ found
solutions of Bianchi type V with an imperfect fluid.
Koppar and Patel$^{25}$ found an exact solution of Bianchi type II 
with viscosity and heat flux.

With these ideas, we write the energy-momentum tensor as
\begin{equation}
{\rm T_{\mu \nu} = \rho\, u_\mu u_\nu + 2q_{(\mu}u_{\nu)} +ph_{\mu\nu} +
\pi_{\mu\nu}},
\end{equation}
where $\rho$ is the energy density of the fluid, $\rm u_\mu$ the velocity,
$\rm q_\mu$ the heat flux vector, p the pressure, $\pi_{\mu\nu}$
the anisotropic stress tensor and
$\rm h_{\mu\nu}=g_{\mu\nu} +u_\mu \, u_\nu$ is the projection orthogonal to
the velocity. Finally, Sec. 5 is devoted to some final remarks.

\section{ Cosmological Model Bianchi type I}
\setcounter{equation}{0}

We start with Bianchi type I metric in five dimensions
\begin{equation}
{\rm ds^2= -dt^2+ e^{2A}\, dx^2 + e^{2B}\, dy^2+ e^{2C}\, dz^2
+ \phi^2\, dw^2,}
\label {2.1}
\end{equation}
where all function A, B, C and $\phi$ depend only to time t. The Einstein
field equations for this model are
\begin{eqnarray}
&&{\rm \dot A \dot B + \dot A \dot C + \dot B \dot C +
(A+B+C\dot ){\dot \phi\over \phi}} = 0, \\
&&{\rm \ddot B +(\dot B)^2 + \dot B \dot C + (\dot C)^2 +\ddot C
 +(B+C \dot ) {\dot \phi\over \phi} + {\ddot \phi\over \phi}} =0, \\
&&{\rm \ddot A+(\dot A)^2 + \dot A \dot C + (\dot C)^2 + \ddot C
 +(A+C \dot ) {\dot \phi\over \phi} + {\ddot \phi\over \phi}} =0, \\
&&{\rm \ddot A +(\dot A)^2 + \dot A \dot B +(\dot B)^2+ \ddot B
 +(A+B \dot ) {\dot \phi\over \phi} + {\ddot \phi\over \phi}} =0, \\
&&{\rm \ddot A +\ddot B + \ddot C +(\dot A)^2 +(\dot B)^2+ (\dot C)^2 +
 \dot A \dot B  + \dot B \dot C+\dot A \dot C  } =0.
\end{eqnarray}
The solution for this set of equations is:

\begin{equation}
\rm \phi (t) = \phi_0 \, t^{k/{(1+k)}}, \qquad \frac{k}{1+k}\not=0,
\label {2.24}
\end{equation}

\begin{eqnarray}
{\rm A} &=& {\rm  \ln \Big (  A_0 \, t^{p_1/{(1+k)}} \Big )},\\
{\rm B} &=& {\rm  \ln \Big (  B_0 \, t^{p_2/{(1+k)}} \Big )},\\
{\rm C} &=& {\rm  \ln \Big (  C_0 \, t^{p_3/{(1+k)}} \Big )},
\label {2.27}
\end{eqnarray}
with $$\rm k\not = 0, \qquad \sum_{i=1}^3 p_i=1, \qquad
 \sum_{i=1}^3 p_i^2 = 1+2k.$$
In order to have a decreasing $\phi (t)$,
it is necessary to have $\rm \frac{k}{1+k} \, < 0$, and using the fact that
$\sum_{i=1}^3 p_i^2 = 1+2k>0$, then $\rm -\frac{1}{2} < k <0$,
therefore when
$t \rightarrow \infty$, $\phi \rightarrow 0$, giving a dynamical
compactification in five dimension.

\section{ Cosmological Model Bianchi type V }
\setcounter{equation}{0}
We begin with the line element
\begin{equation}
{\rm  ds^2 = - e^{2A}d\tau^2 + e^{2A} dx^2+ e^{2B-2Qx} dy^2 + e^{2C-2Qx}
dz^2 + e^{ 2\psi} d\omega^2}, \label {3.1}
\end{equation}
which defines the anisotropic and homogeneus cosmological model Bianchi V,
where $\rm A, B, C$ and $\psi$ are functions of the ``time parameter''
$\tau$, $\rm Q$ is a constant parameter that
give us a measure of the anisotropy of this model,
$g_{\scriptstyle 44}= e^{2\psi}$,
where $\phi=e^\psi$ is a massles scalar field associated to the extra fifth
dimension $x^4 = \omega$.

The five-dimensional vacuum field equations for this model are:
\begin{eqnarray}
 &&{\rm 2 A'} ={\rm B' + C'}, \\
 &&{\rm A' B' + A' C' + B' C' - 3 Q^2} = -{\rm  (A'+B'+C') \psi'},\\
 &&{\rm B'' + B'^2 + B' C'+ C'^2 + C'' - A'(B'+C') - Q^2} = \nonumber\\
 & &\mbox{} - {\rm \bigg ((B'+C'-A')\psi'+\psi'^2+\psi''\bigg)},\\
&& {\rm A'' + C'' + C'^2 - Q^2} =
- {\rm \bigg ( C' \psi'  + \psi'^2 + \psi'' \bigg )},\\
&&{\rm  A'' + B'' + B'^2 - Q^2} =
- {\rm \bigg ( B' \psi'  + \psi'^2 + \psi'' \bigg )},\\
&& {\rm A'' + B'' + C'' + B'^2 + C'^2 + B' C' - 3Q^2 }= 0,
\end{eqnarray}
where now, the primes mean derivatives with respect to the ``time variable
$\tau$''.
The above system is a very symmetric one, and in order to solve it, we
took advantage of this fact. The solutions for the above set of equations
are
\begin{equation}
{\rm  \psi(\tau) = \psi_0 + {1\over {4 Q \alpha_1 \gamma}}
\ln \bigg[ { e^{2Q\tau}-\gamma \over { e^{2Q\tau}+\gamma}} \bigg]},
\label {3.25}
\end{equation}
then
\begin{equation}
{\rm  \phi(\tau) = \phi_0 \bigg[ { e^{2Q\tau}-\gamma \over
{ e^{2Q\tau}+\gamma} } \bigg]^{1\over {(4 Q \alpha_1 \gamma)} } },
\label {3.251}
\end{equation}
where $\psi_0$ and $\gamma$  are constants of integration.
\begin{equation}
{\rm  A= ln \Big [A_0 \Big \lbrace \Big (
{e^{2Q\tau} - \gamma \over {e^{2Q\tau} + \gamma}} \Big
)^{1\over 2Q\gamma\alpha_1}
\Big( {1\over { \alpha_1 e^{2Q\tau} + \alpha_2 e^{-2Q\tau}}} \Big )
\Big \rbrace^{-{1\over 2}} \Big ]},  \label {3.26}
\end{equation}
\begin{equation}
\rm B= A + {1\over 2} \alpha_3 \psi + \alpha _4 , \label {3.36}
\end{equation}
\begin{equation}
\rm C= A - {1\over 2} \alpha_3 \psi + \alpha _5 , \label {3.37}
\end{equation}
where $\rm \alpha_4, \alpha_5$ are integration constants and
the following relation is satisfied
\begin{equation}
\rm \alpha_1 \alpha_2 = - {{3+\alpha_3^2}\over {48Q^2}}. \label {3.38}
\end{equation}

Using $\rm d\tau=e^{-A} dt$ in the time reparametrization t, we can get
the following
\begin{equation}
\rm  t-t_0 = \int_{\tau_0}^{\tau} \Big [A_0 \Big / \Big \lbrace \Big (
{{e^{2Q\tau} - \gamma}\over {e^{2Q\tau} + \gamma} }
 \Big )^{1\over 2Q\alpha_1\gamma}
\Big( {1\over { \alpha_1 e^{2Q\tau} + \alpha_2 e^{-2Q\tau}}} \Big )
\Big \rbrace^{1\over 2} \Big ] d\tau.
\label {3.39}
 \end{equation}
It is reasonable to think that the time $\tau$ has a behavior similar
to time t,
because, when $\tau \rightarrow \infty$, the integral goes to $\infty$ too.
On the other hand, taking a look to Eq. (3.9), when
$\tau \rightarrow \infty$, the function $\phi \rightarrow \phi_0=\rm const.$,
to have dynamical compactification we chose a small value for $\rm \phi_0$.
Moreover, as can be seen from Eqs. (3.10-3.12),
our model isotropize for $\tau \rightarrow \infty$.

\section{ Stress-energy Tensor in four dimensions}
\setcounter{equation}{0}
In the set of equation (2.2) to (2.6) for the Bianchi type I, the terms
in which the scalar field $\phi$ appears can be associated to the components
of an imperfect fluid stress-energy tensor in four dimensions in the
following sense,
\begin{eqnarray}
{\rm -(A+B+C\dot ){\dot \phi\over \phi}} &=&8\pi \rho= T^0_0,\\
{\rm (B+C \dot ) {\dot \phi\over \phi} + {\ddot \phi\over \phi}} &=&
8\pi p_1 = T^1_1,\\
{\rm (A+C \dot ) {\dot \phi\over \phi} + {\ddot \phi\over \phi}} &=&
8\pi p_2=T^2_2,\\
{\rm (A+B \dot ) {\dot \phi\over \phi} + {\ddot \phi\over \phi}} &=&
8\pi p_3=T^3_3.
\end{eqnarray}
In a similar way, for the Bianchi type V, Eqs. (3.3) to (3.6), become
\begin{eqnarray}
- {\rm (A+B+C \dot )\dot \psi}&=& 8\pi \rho=T^0_0,\\
 {\rm  (B+C \dot ) \dot \psi  + {\dot \psi}^2 + \ddot \psi}
&=&8\pi p_1=T^1_1 ,\\
{\rm  (A+C \dot ) \dot \psi  + {\dot \psi}^2 + \ddot \psi }
&=&8\pi p_2=T^2_2, \\
 {\rm  (A+B \dot ) \dot \psi  + {\dot \psi}^2 + \ddot \psi }
&=&8\pi p_3=T^3_3.
\end{eqnarray}
For these models, in our case, if  we identify Eqs. (2.2-2.5) and (3.3-3.6)
in four
dimension with a matter
tensor $\rm T_{\mu\nu} $,  the components of this tensor obey the relation
$\rm  T^\mu_\mu = 0$ when we use the constraints equations (2.6) and (3.7),
which imply that the matter is ultrarelativistic or radiation
like. The anisotropy represented by $ T^1_1 \ne T^2_2 \ne T^3_3 $ is due
physically to the inequality of the pressures. Thus, although the fluid
(scalar fied) doesn't have viscosity it has an energy-momentum tensor that
differs from the conventional one
\begin{equation}
\rm   T_{\mu\nu} = [(p + \rho)u_\mu u_\nu + p g_{\mu\nu}],
\end{equation}
for perfect fluids. However, in those cases we can derive an equation of
state by using the  technique of  the references [3,8] in
which they identify
$\rm - T^0_0 = \rho_{eff}$ and $\rm {1\over 3} T^i_i=p_{eff}$, in order to
have $\rm \rho_{eff} = 3 p_{eff}$.
In the following subsection we show that the analogy with an imperfect fluid
is more general.
\subsection { Structure of the imperfect fluid Energy-Momentum Tensor }

We recall here the
 explicit field equations for this theory,
\begin{equation}
{\rm G_{\mu \nu}}={\rm T_{\mu \nu}(\phi):=\phi^{-1}(\phi_{;\mu\nu}-
g_{\mu\nu} \square \phi)}.
\label {a}
\end{equation}
\begin{equation}
\square \phi=0,
\label {b}
\end{equation}
where $\rm G_{\mu \nu}$ is the Einstein tensor .

The defined energy momentum for the scalar field is covariantly conserved as
follows from the Bianchi identities or the field equation for $\phi $,
\begin{equation}
{\nabla}_{\mu} T^{\mu \nu}(\phi )=0.
\label  {3}
\end{equation}

We will show that this energy momentum tensor has the structure of an
imperfect fluid energy-momentum tensor
\begin{equation}
T_{\mu \nu} = \rho u_{\mu} u_{\nu}+ 2 q_{(\mu} u_{\nu)}+ p h_{\mu \nu}
+\pi_{\mu \nu},
\label {4}
\end{equation}
where $\rho$ is the energy density of the fluid, $u_{\mu}$ the
velocity, $q_{\mu}$ the heat flux, p the pressure, $\pi_{\mu \nu}$ the
anisotropic stress tensor and,
\begin{equation}
h_{\mu \nu}= g_{\mu \nu }+ u_{\mu} u_{\nu}
\label {5}
\end{equation}
is the projection orthogonal to the velocity. The following relations are
satisfied,
\begin{equation}
 u_{\mu} u^{\mu}= -1;\quad h_{\mu \nu} u^{\mu} =0;\quad \pi_{\mu}^{ \mu}=0;
\quad h_{\mu \nu}  u^{\mu} u^{\nu}=0.
\label {6}
\end{equation}

The thermodynamic quantities can be obtained by projections along and
orthogonal to the velocity field and taking traces,
\begin{equation}
\rho = T_{\mu \nu}u^\mu u^\nu ,\quad q_{\alpha}=-T_{\mu \nu} u^{\mu}
h^{\nu}_{\alpha},
\end{equation}
\begin{equation}
 \Pi_{\alpha \beta}:= p h_{\alpha \beta}+ \pi_{\alpha \beta}= -T_{\mu \nu}
 h^{\mu}_{\alpha} h^{\nu}_{\beta},
\end{equation}
\begin{equation}
 p= {1\over 3} \Pi^{\alpha}_{\alpha},
\end{equation}
\begin{equation}
 \pi _{\alpha \beta} =\Pi _{\alpha \beta}- p h_{\alpha \beta}.
\label {7}
\end{equation}
Also of interest are the kinematical quantities of the fluid, that appear in
the following decomposition of the derivative of the velocity
\begin{equation}
u_{\alpha ;\beta }= -{\dot u_\alpha }u_\beta  + \omega_{\alpha \beta }+
 \sigma _{\alpha \beta }+\theta h _{\alpha \beta }/3,
\label {8}
\end{equation}
with
\begin{equation}
{\dot u}_{\alpha}= u_{\alpha ;\beta } u^{\beta },
\end{equation}
\begin{equation}
 \theta = u_{; \alpha }^{;\alpha },
\end{equation}
\begin{equation}
 \sigma_{\alpha \beta }
  =u_{(\alpha ;\beta) }+{\dot u}_{(\alpha} u_{\beta )}-{1
\over 3} \theta h_{\alpha \beta },
\end{equation}
\begin{equation}
\omega_{\alpha \beta }= u_{[\alpha ;\beta] }+{\dot u}_{[\alpha}
u_{\beta ]},
\label {9}
\end{equation}
which are the acceleration, expansion, shear and rotation of the fluid
respectively.  In this case we take for the velocity of the fluid the
normalized derivative of the field,
\begin{equation}
u_{\alpha} := \pm {{\phi_{;\alpha}}\over {\sqrt {Z}}}
\label {10}
\end{equation}
where $ Z:=-\phi _{;\sigma}\phi^{;\sigma}$ and we have to choose the most
convenient sign in each particular calculation. In order to make sense for
the last expression it is necessary that $\phi _{;\sigma}$ be a timelike
vector;
this condition could be a serious restriction to the aplicability of the
generating method. We shall see latter that there are situations of physical
interest where the condition is fullfilled.

Having selected the velocity field we use it in Eqs. (\ref {7}, \ref{9})
 to obtain the
thermodynamic and kinematical quantities of the fluid in terms of the scalar
 field and its derivatives. After a straightforward calculation we obtain
\begin{equation}
\rho=   \frac{Y}{ Z \phi},  \label {11}
\end{equation}

\begin{equation}
p=\frac{1}{3} \frac{Y}{ Z \phi},
\label {12}
\end{equation}

\begin{equation}
q_\alpha =\mp Z^{-{1\over 2}}\phi^{-1}\bigl [\phi^{;\nu}\phi_{;\nu \alpha}+
 {{Y \phi_{;\alpha}}\over Z}\bigr ],
\label {13}
\end{equation}

\begin{equation}
\Pi_{\alpha \beta}=
{{\phi_{;\alpha}\phi_{;\beta}}
\over Z}\bigl [  { Y  \over { Z \phi}}\bigr ]+{\phi_{;\alpha \beta} \over 
{\phi}}
+2{\phi_{(;\alpha}{\phi_{;\mu \beta)}  \phi^{;\mu}} \over { Z \phi}},
\label {14}
\end{equation}

\begin{equation}
\pi_{\alpha \beta}=-\frac{1}{3} {g_{\alpha \beta}} \bigl [
{ Y \over { Z \phi}} \bigr ]
+{{ \phi_{;\alpha ;\beta}}\over \phi }
+{1 \over 3}{{\phi_{;\alpha}\phi_{;\beta}}\over \phi}\bigl [
 2 {Y  \over { Z^2}}\bigr ]
+2{\phi_{(;\alpha}{\phi_{;\mu \beta)}  \phi^{;\mu}} \over { Z \phi}} ,
\label {15}
\end{equation}

\begin{equation}
{\dot u}_{\alpha } ={ {\phi_{;\alpha ;\beta } \phi^{;\beta }} \over Z} +
{{Y{ \phi_{;\alpha }} \over Z^2}} ,
\label {16}
\end{equation}

\begin{equation}
\theta =\pm \bigl [ {Y \over Z^{3 \over 2}}
	  \bigr ] ,
\label {17}
\end{equation}

\begin{equation}
\omega_{\alpha \beta }=0,
\label {18}
\end{equation}

\begin{equation}
 \sigma _{\alpha \beta }=\pm \bigl \{
 -{1\over 3}{{g_{\alpha \beta}}}
     \bigl [ { Y \over { Z ^{3/2}}} \bigr ]
+{{ \phi_{;\alpha ;\beta}}\over Z^{1/2} }
+{1\over 3}\phi_{;\alpha } \phi _{;\beta }
   \bigl [{2Y\over Z^{5 \over 2}}\bigr ]
+2{\phi_{(;\alpha}{\phi_{;\mu \beta)}  \phi^{;\mu}} \over
     { Z^{3\over 2} }}\bigr \},
\label {19}
\end{equation}
 here
\begin{equation}
Y := \phi _{;\lambda \mu} \phi ^{;\lambda} \phi ^{;\mu} .
\label {20}
\end{equation}

We notice  that the fluid that we have obtained is irrotational because the
velocity field is propotional to a divergence, it is also a Newtonian fluid,
i.e., satisfying the relation
\begin{equation}
\pi _{\alpha \beta} = -2\; \eta\; \sigma_ {\alpha \beta},
\label {21a}
\end{equation}
 with the viscosity coefficient given by
\begin{equation}
\eta = \mp {Z^{1 \over 2}\over {2\phi}},
\label {21b}
\end{equation}
also the equation of state is $\rm p=\frac{1}{3} \rho$ that correspond
to a relativistic fluid, but an imperfect one since heath flux and viscosities
are present.

\subsection { Examples}

In this section we apply the results obtained above to generate solutions in
the ani\-so\-tro\-pic homogeneous cosmology. In this models the scalar field 
depends
only on the time coor\-di\-nate and the derivative is a timelike vector. We
exhibit the thermodynamic and kinematical quantities obtained for the
solutions of Bianchi type I and Bianchi type V.

\subsubsection {Bianchi type I solution}.

Here we consider the solutions (\ref {2.24}-\ref{2.27}),
substituting these solutions in Eqs.(4.26 - 4.37)
we obtain the following thermodynamic and kinematical quantities:
\begin{equation}
\rm \rho =- \frac{k}{(1+k)^2\,t^2},
\end{equation}

\begin{equation}
\rm  p=- \frac{1}{3} \, \frac{k}{(1+k)^2 \, t^2},
\end{equation}

\begin{equation}
\rm  q_\alpha =0,
\end{equation}

\begin{equation}
\rm \pi_{ij} =- \frac{k}{6(1+k)^2} \Big(3p_i-2 \Big) \,
t^{[\frac{p_i}{1+k}-2]} \qquad \delta_{ij},
\end{equation}

\begin{equation}
\rm \eta = - \frac{k}{2(1+k)\, t^2} ,
\end{equation}

\begin{equation}
\rm \theta = \frac{1}{(1+k) \, t}.
\label {24}
\end{equation}

 In order to have an expanding universe ($\theta >0$)and a positive
viscosity coefficient we have to put the restriction
$-1<k< 0$, being this to agreement with the limits
set at the end of Sec. 2, and also the pressure
and energy density are positive.

In this particular case the fluid is not only irrotational and
Newtonian, the viscosity is also proportional to the energy density and the
equation of state is of the barotropic type.

\subsubsection {Bianchi type V solution}.

Here we consider the solutions (\ref {3.25}-\ref{3.37}),
we obtain the following thermodynamic and kinematical quantities for this
model
\begin{equation}
\rm \rho = \frac{3}{2} \frac{e^{4Q\tau} \lbrace
\frac{e^{2Q\tau}-\gamma}{e^{2Q\tau}+\gamma}
   \rbrace^{1/(4\alpha_1 Q \gamma)} \Big ( e^{2Q \tau} - 2\alpha_1 Q
e^{4Q \tau}-2\alpha_1 \gamma^2 Q \Big) }{\alpha_1^2 (\alpha_2+\alpha_1
e^{4Q\tau})(e^{4Q\tau} - \gamma^2)^2 },
\end{equation}

\begin{equation}
\rm  p= \frac{1}{2} \frac{ e^{4Q\tau} \lbrace
\frac{e^{2Q\tau}-\gamma}{e^{2Q\tau}+\gamma}
   \rbrace^{1/(4\alpha_1Q \gamma)} \Big ( e^{2Q \tau} - 2\alpha_1 Q e^{4Q\tau}
-2\alpha_1 \gamma^2 Q \Big) }{\alpha_1^2 (\alpha_2+\alpha_1 e^{4Q\tau})
(e^{4Q\tau} - \gamma^2)^2 },
\end{equation}

\begin{equation}
\rm  q_\alpha =0,
\end{equation}

\begin{eqnarray}
\pi_{11}&=&\frac{1}{2} \frac{1-2Q\Big(\alpha_1e^{2Q\tau}-\alpha_2e^{-2Q\tau}
\Big)}{(\alpha_1 e^{2Q\tau}+\alpha_2e^{-2Q\tau})^2}, \nonumber\\
\pi_{22}&=& \frac{1}{2} \frac{1-2Q\Big(\alpha_1e^{2Q\tau}-\alpha_2e^{-2Q\tau}
\Big)-\alpha_3}{(\alpha_1 e^{2Q\tau}+\alpha_2e^{-2Q\tau})^2}
e^{-2Qx} \phi^{\alpha_3},\nonumber\\
\pi_{33}&=& \frac{1}{2} \frac{1-2Q\Big(\alpha_1e^{2Q\tau}-\alpha_2e^{-2Q\tau}
\Big)+ \alpha_3}{(\alpha_1 e^{2Q\tau}+\alpha_2e^{-2Q\tau})^2}
e^{-2Qx} \phi^{-\alpha_3},
\end{eqnarray}

\begin{equation}
\rm \eta =  \frac{1}{2} \Big ( \frac{e^{2Q\tau}-\gamma}{e^{2Q\tau}+\gamma}
\Big)^{1/(4Q\alpha_1 \gamma)} \Big( \frac{1}{\alpha_1 e^{2Q\tau}
 +\alpha_2 e^{-2Q\tau}} \Big)^{\frac{3}{2}} ,
\end{equation}

\begin{equation}
\rm \theta =- 3 \frac{e^{Q\tau}  \Big ( e^{2Q \tau} - 2\alpha_1 Q
e^{4Q \tau}-2\alpha_1 \gamma^2 Q \Big) }{\alpha_1^{3/2}
(e^{4Q\tau} - \gamma^2)^{3/2} },
\end{equation}


\section {Final remarks}
In this work we have considered  five dimensional cosmological models
that have as four dimensional hypersurface Bianchi type I and V
homogeneous spaces. We have found exact solutions where the influence of
the extra dimension is encoded in a scalar field that can be considered
the radius of that extra dimension. The model that we found are such
that for later times the scalar field either vanishes or take a constant
very value that can be chosen small so that we have dynamical
compactification of the fifth dimension. The Bianchi type V also
isotropize at later times. On the other hand, using the standard dimensional
reduction, the energy momentum tensor of the scalar field can be
associated with an imperfect fluid that has dissipative effects: heat
flux and shear viscosity that can be useful to explain several features
of the evolution of the universe as entropy generation. The relation
between the pressure and the energy density corresponds to a
relativistic fluid, $p=\rho/3$. The fluid is also Newtonian.
\bi\bi

\noi {\Large Acknowledgments}

J.S. and V.M.V. were supported in part by CONACYT grant 4862-E9406, L.O.P.
was supported in part by CONACYT grant 1861-E9212.

\newpage
{\Large References}
\begin{enumerate}

\item  Des J. Mc Manus, J. Math. Phys. {\bf 35}, 4889 (1994).
\item  Chodos  and S. Detweiler,   Phys. Rev. D {\bf 21}, 2167
	(1980).
\item  A. Davidson  and D.A. Owen, Phys. Lett. B {\bf 155}, 247
	 (1985).
\item P.S. Wesson,  Phys. Lett. B {\bf 276}, 299 (1992).
\item  A. Einstein, {\it The Meaning of Relativity} (Princeton 1956).
\item R. Balbinot, J. Fabris and R. Kerner, Phys. Rev. D
	 {\bf 42}, 1023 (1990).
\item  Ponce de Leon, Gen. Rel. Grav.{\bf 20}, 539 (1988).
\item P.S. Wesson,  Astrophys. J. {\bf 394}, 2167 (1992).
\item  P.G.O. Freund, Nucl. Phys. B {\bf 209}, 146 (1982)
\item  E. Alvarez and M.B. Gavela, Phys. Rev. Lett. {\bf 51}, 931 (1983).
\item  D. Sahdev, Phys. Lett. B {\bf 137}, 155 (1984).
\item R.B. Abbott, S.M. Barr  and S.D. Ellis, Phys. Rev. D
	{\bf 30}, 720 (1984).
\item E.W. Kolb, D.L. Lindley and D. Seckel, Phys. Rev.
	D {\bf 30}, 1209 (1984).
\item D. Lorenz-Petzold, Phys. Lett. B {\bf 153}, 134 (1985).
\item L. O. Pimentel, Class. Quantum Grav. {\bf 9}, 377 (1992).
\item L. O. Pimentel and J. Socorro, Int. J. Math. Phys. {\bf 34},
	701 (1995).
\item G. L. Murphy, Phys. Rev. D {\bf 8}, 4231 (1973).
\item A. Banerjee and N. O. Santos, J. Math. Phys. {\bf 26},
	878 (1985).
\item V. A. Belinskii and I. M. Khalatnikov, Sov. Phys. JETP,
 {\bf{45}}, 1 (1977).
\item V. A. Belinskii and I. M. Khalatnikov, Sov. Phys. JETP,
 {\bf{42}}, 2051 (1976).
\item A. Banerjee, S.B. Duttachoudhury and A.K. Sanyal,
  J. Math. Phys. {\bf 26}, 3010 (1985).
 \item A. Banerjee, S.B. Duttachoudhury and A.K. Sanyal,
 Gen. Rel.Grav. {\bf 18}, 461 (1986).
\item  J.M. Bradley and E. Sviestins, Gen. Rel. Grav. {\bf 16},
	1119 (1984).
\item  B.K. Nayak and B.K. Sahoo, Gen. Rel. Grav. {\bf 21},
	211 (1989).
\item  S.S. Koppar and L.K. Patel, Acta Physica
	Hungarica. {\bf 67}, 297 (1990).

\end{enumerate}
\end{document}